\RequirePackage[T1]{fontenc}
\documentclass[twoside,12pt]{article}
\usepackage{times,amsmath,amsfonts,amsthm,amssymb,eucal}
\usepackage{bbold}
\usepackage{graphicx,ifthen}
\usepackage{wrapfig}             
\usepackage{latexsym}
\usepackage{color}
\usepackage{mathrsfs}
\usepackage{bm}
\usepackage{algorithmicx}
\usepackage{algorithm}
\usepackage{algpseudocode}
\usepackage{rotating}

\usepackage{booktabs}
\usepackage{multirow}

\usepackage{hyperref}

\usepackage{subcaption}

\begin{document}
\DeclareGraphicsExtensions{.gif,.pdf,.png,.jpg,.tiff}

\def\({\left(}
\def\){\right)}
\def\[{\left[}
\def\]{\right]}
\def\a{\alpha}\def\b{\beta}\def\d{\delta}\def\D{\Delta}
\def\e{\epsilon}\def\g{\gamma}\def\G{\Gamma}\def\k{\kappa}
\def\l{\lambda}\def\L{\Lambda}\def\m{\mu}\def\p{\phi}\def\P{\Phi}
\def\r{\rho}\def\s{\sigma}\def\t{\theta}\def\T{\Theta}
\def\ta{\tau}\def\z{\zeta}
\def\ralph{\mathscr} 
\def\aa{{\ralph A}}\def\bb{{\ralph B}}\def\cc{{\ralph C}}
\def\dd{{\ralph D}} \def\ee{{\ralph E}}\def\ff{{\ralph F}}\def\gg{{\ralph G}}
\def\hh{{\ralph H}} \def\ii{{\ralph I}}\def\jj{{\ralph J}}\def\kk{{\ralph K}}
\def\ll{{\ralph L}} \def\mm{{\ralph M}}\def\nn{{\ralph N}}\def\oo{{\ralph O}}
\def\pp{{\ralph P}} \def\qq{{\ralph Q}}\def\rr{{\ralph R}}\def\ss{{\ralph S}}
\def\tt{{\ralph T}} \def\uu{{\ralph U}}\def\vv{{\ralph V}}\def\ww{{\ralph W}}
\def\xx{{\ralph X}} \def\yy{{\ralph Y}}\def\zz{{\ralph Z}}
\font\Caps=cmcsc10
\font\BigCaps=cmcsc9 scaled \magstep 1
\font\BigSlant=cmsl10    scaled \magstep 1
\font\proclaimfont=cmbx9 scaled \magstep 1
\def\BigHeading{\bfseries\Large}\def\MediumHeading{\bfseries\large}
\def\smc{\Caps}
\def\lbk{\linebreak}
\newdimen\bigindent
\newdimen\smallindent
\bigindent=30pt
\smallindent=5pt
\def\quoteindent{\advance\leftskip by\bigindent\advance\rightskip
                 by\bigindent}
\newskip\proclaimskipamount
\proclaimskipamount=12pt  plus1pt minus1pt
\def\proclaimskip{%
  \par\ifdim\lastskip<\proclaimskipamount
  \removelastskip\vskip\proclaimskipamount\fi}
\let\demoskip=\proclaimskip
\def\Demo#1{\par\ifdim\lastskip<\proclaimskipamount
            \removelastskip\proclaimskip\fi
        \sl#1. \hskip\smallindent\rm}
\def\EndDemo{\par\demoskip}
\def\DemoSection#1{\par\ifdim\lastskip<\proclaimskipamount
             \removelastskip\proclaimskip\fi
             #1\hskip\smallindent\rm}
\def\Section#1{\stepcounter{section}
    \DemoSection{{\bfseries\large\thesection.\hskip\smallindent#1.}}}
\def\Subsection#1{\stepcounter{subsection}
    \DemoSection{\bfseries\normalsize\thesubsection.\hskip\smallindent#1.}}
\def\Quote{\begin{quotation}\normalfont\small}
\def\EndQuote{\end{quotation}\rm}

\newtheorem{theorem}{Theorem}
\newtheorem{lemma}{Lemma}
\newtheorem{corollary}{Corollary}
\newtheorem{remark}{Remark}
\newtheorem{example}{Example}
\newtheorem{definition}{Definition}
\newtheorem{assumption}{Assumption}
\newtheorem{strategy}{Strategy}

\def\Theorem{\begin{theorem}\sl}
\def\EndTheorem{\end{theorem}}
\def\Lemma{\begin{lemma}\sl}
\def\EndLemma{\end{lemma}}
\def\Corollary{\begin{corollary}\sl}
\def\EndCorollary{\end{corollary}}
\def\Remark{\begin{remark}\rm}
\def\EndRemark{\end{remark}}
\def\Definition{\begin{definition}\sl}
\def\EndDefinition{\end{definition}}
\def\Assumption{\begin{assumption}\sl}
\def\EndAssumption{\end{assumption}}
\def\bct{\begin{center}}
\def\ect{\end{center}}
\def\Array{\begin{eqnarray*}}
\def\EndArray{\end{eqnarray*}}
\def\Enumerate{\begin{enumerate}}
\def\EndEnumerate{\end{enumerate}}
\def\Eq{\begin{equation}}
\def\EndEq{\end{equation}}
\def\EqArray{\begin{eqnarray}}
\def\EndEqArray{\end{eqnarray}}
\def\mref#1{(\ref{#1})}
\def\qt#1{\qquad\text{#1}}
\def\Tabular{\begin{tabular}}
\def\EndTabular{\end{tabular}}
\def\FlushLeft{\begin{flushleft}}
\def\EndFlushLeft{\end{flushleft}}
\newcommand{\red}{\textcolor{red}}
\newcommand{\blue}{\textcolor{blue}}
\newcommand{\green}{\textcolor{green}}
\def\pmb#1{\setbox0=\hbox{#1}%
 \kern-0.010em\copy0\kern-\wd0
 \kern0.035em\copy0\kern-\wd0
 \kern0.010em\raise.0233em\box0}
\def\confirm{\mathrm{confirm}}
\def\death{\mathrm{death}}
\def\imp{\mathrm{imp}}
\def\pop{\mathrm{pop}}
\def\erf{\operatorname{erf}}
\def\N{\text{N}}
\def\X{{\bf X}}
\def\x{{\bf x}}
\def\Y{{\bf Y}}
\newcommand{\oob}[1][i]{\ensuremath{\mathcal{O}_{#1}}}
\def\RFSRC{{\ttfamily randomForestSRC}}
\def\cmprsk{{\ttfamily cmprsk}}
\def\crrstep{{\ttfamily crrstep}}
\def\riskregression{{\ttfamily riskRegression}}
\def\CoxBoost{{\ttfamily CoxBoost}}

\def\E{\mathbb{E}}
\def\PP{\mathbb{P}}
\def\QQ{\mathbb{Q}}


\def\Report{Dynamic Competing Risk Modeling COVID-19}
\def\Author{Min Lu and Hemant Ishwaran}
\pagestyle{myheadings}\markboth{\Author}{\Report}
\thispagestyle{empty}

{
  \title{\bf Dynamic Competing Risk Modeling COVID-19 \\
    in a Pandemic Scenario}
  \author{Min Lu\thanks{Supported by National Institutes
  		Health grants R01 CA200987 and R01 HL141892.}   $\,$and Hemant Ishwaran\thanks{Supported by National Institutes
      Health grants R01 GM125072 and R01 HL141892.}
  \\
  Department of Public Health Sciences, University of Miami}
  \maketitle
  }

\Quote\vskip-30pt\noindent The emergence of coronavirus disease 2019
(COVID-19) in the United States has forced federal and local
governments to implement containment measures. Moreover,  the severity of the situation
has sparked engagement by both the research and clinical community
with the goal of developing effective treatments for the disease. This
article proposes a time dynamic prediction model with competing risks
for the infected individual and develops a simple tool for policy
makers to compare different strategies in terms of when to implement
the strictest containment measures and how different treatments can
increase or suppress infected cases. Two types of containment
strategies are compared: (1) a constant containment strategy that
could satisfy the needs of citizens for a long period; and (2) an
adaptive containment strategy whose strict level changes across
time. We consider how an effective treatment of the disease can affect
the dynamics in a pandemic scenario.  For illustration we consider a
region with population 2.8 million and 200 initial infectious cases
assuming a 4\% mortality rate compared with a 2\% mortality rate if a
new drug is available.  Our results show compared with a constant
containment strategy, adaptive containment strategies shorten the
outbreak length and reduce maximum daily number of cases. This, along
with an effective treatment plan for the disease can minimize death
rate.  \EndQuote

\noindent%

\thispagestyle{empty}

\section{Introduction}
\label{sec:intro}

To prevent the spread of a new infectious disease such as coronavirus
disease 2019 (COVID-19), policy makers rely on prediction models to
foresee the number of infectious cases and to inform best 
containment measure strategies including patient quarantine, active monitoring
of contacts, border controls, and community education and
precautions~\cite{shearer2020infectious,ng2020evaluation,hunter2020covid,kupferschmidt2020will}. There
are many prediction models available for this kind of 
modeling~\cite{dye2003modeling,bauch2005dynamically,huang2004simulating,colizza2007modeling,rahmandad2008heterogeneity,gray2011stochastic,capasso1978generalization,capasso2008mathematical,liu1986influence,zhang2005compartmental}.
In predicting local COVID-19 spread, there are two major
challenges. Firstly, number of actual infected cases is usually
unconfirmed and could be far larger than confirmed cases because there
are significant number of infected cases in incubation period and test
kits may be insufficient. Secondly, regions that experienced
earlier outbreaks can provide valuable information, such as the
distribution of cure time, death time, and mortality
rate~\cite{wilson1994analysis}, but it is not easy to integrate these
dynamic parameters into many current models.

This article provides a simple and robust model framework whose
parameters are dynamically adjustable and generally interpretable for
policy makers.  This framework utilizes competing risks survival
analysis to borrow information from regions that experienced earlier
outbreaks.  Moreover, the model enables containment measures to change
over time~\cite{cohen2020strategies} through introducing a novel
transmission number which incorporates containment measures and the
basic reproduction number ($R_0$).

\section{The model}
\label{section:model}

Assume the disease of interest has a $M$-day period of communicability
so that infected people are either cured or dead within $M$ days. The
value $M$ can also be treated as a parameter in our model.  Denote the
mortality rate within an infectious period as $m_{\death}$. On day
$t$, denote the number of people that have been infected for $d$ days
as $p_{t,d}$.  The total number of infectious cases at time $t$ is
$P_t=\sum_{d=1}^{M}p_{t,d}$, where $p_{t,d}$ is determined by the
following factors:

\Enumerate\setlength{\itemsep}{0pt}
\item Mortality rate for people that have been infected for $d$ days, denoted as $m_d$.
\item Cure rate for people that have been infected for $d$ days, denoted as $c_d$.
\item Average number of people an infectious person can communicate on day $t$, denoted as $R_t$.
\item Number of travelers from other areas who have been infected for $d$ days, denoted as $p^{\imp}_{t,d}$.
\EndEnumerate

When moving forward from day $t$ to $t+1$, the number of infectious
cases, $P_{t+1}$, is the sum of three terms: (a) the number of
survived but uncured cases from day $t$; (b) the number of newly
infected cases; and (c) the number of imported cases, denoted as
$P^{\text{imp}}_{t+1}=\sum_{d=1}^{M}p^{\imp}_{t,d}$~\cite{chinazzi2020effect,layne2020new,pacheco2020dispersion}:
\Eq
P_{t+1} :=\sum_{d=1}^M p_{t+1,d} = \sum_{d=1}^{M-1}
p_{t,d}(1-m_d-c_d) + P_t R_t + P_{t+1}^{\imp}.
\label{dynamic.model}
\EndEq
Here we use $p_{t+1,1}=P_tR_t$, which counts newly infected cases, and
for $d=1,\dots,M-1$, we have $p_{t+1,d+1}=p_{t,d}(1-m_d-c_d)$.  Note
that the people who have been infected for $M$ days on day $t$
($p_{t,M}$) will not affect $P_{t+1}$ since their period of
communicability will be over and they will be either dead or cured on
day $t+1$.

\section{Competing risk survival analysis for mortality and cure parameter specification}
\label{section:para.cr}

We use a competing risks framework to specify the mortality rate $m_d$
and cure rate $c_d$.  Let $T$ be the continuous event time of an
infected patient.  Notice that $T$ is subject to two mutually exclusive
competing risks: cure or death.  Let $\d\in\{1,2\}$ be the indicator
recording which event occurs;  $\d=1$ denotes cure and $\d=2$
denotes death.

The cumulative incidence (CIF) is the probability of experiencing an
event of type $j$ by time $t$, i.e. $F_j(t) = \PP\{T \le t, \d^o=j\}$.
The CIF is related to the survival function $S(t)=\PP\{T\ge t\}$ by the
identity
\Array
S(t)
&=&  1 - \PP\{T\le t\}\\
&=&  1 - \[\PP\{T\le t, \d=1\} +  \PP\{T\le t, \d=2\}\]\\
&=&  1 - F_1(t) - F_2(t).
\EndArray
The cause-specific hazard $h_j$ for event $j$ is given by
$$
h_j(t) = \lim_{\D t\rightarrow 0} \frac{\PP\{t\le T \le t+\D t,
  \d=j | T\ge t\}}{\D t} = \frac{f_j(t)}{S(t)},\qt{}t>0.
$$
Thus $h_j$ has the following intuitive meaning
$$
S(t)h_j(t) \asymp \frac{\PP\{t\le T \le t+\D t, \d=j \}}{\D t}.
$$
From this, one can deduce that
$$
F_j(t) 
=\int_0^t S(s)h_j(s)\, d s
=\int_0^t S(s)\, dH_j(s)
$$
where $H_j(t)=\int_0^t h_j(s)ds$ is the cumulative hazard function (CHF).
By the mutual exclusiveness of the two events, the hazard for $T$ is
$h(t)=h_1(t)+h_2(t)$. Because $T$ is a continuous random
variable, $S(t)=\exp(-H(t))$ where $H(t)=\int_0^s h(s)ds$ is the
CHF.  It follows that 
\Eq
F_j(t) 
= \int_0^t \exp\left(-\int_0^s \sum_{l=1}^2 h_l(u) d u\right) dH_j(s)
= \int_0^t \exp(-H_1(s))\exp(-H_2(s))\, dH_j(s).
\label{cif}
\EndEq
Let $T_j$ be a continuous random variable with hazard $h_j$.  Keep in
mind $T_j$ is used only for theoretical construction and is not
related to $T$.  Let $f_{T_j}$ and $F_{T_j}$ be the density and
cumulative distribution function (CDF) for $T_j$.  Thus
$$
h_j(t)
=\frac{f_{T_j}(t)}{1-F_{T_j}(t)}
=\frac{f_{T_j}(t)}{S_{T_j}(t)}
$$
where $S_{T_j}(t)=\exp(-H_j(t))$ is the survival function for $T_j$.  
Using~\mref{cif}, we can rewrite the CIF as
$$
F_j(t)
= \int_0^t S_{T_1}(s)S_{T_2}(s) h_j(s)\,ds
= \int_0^t S_{T_1}(s)S_{T_2}(s) \frac{f_{T_j}(s)}{S_{T_j}(s)}\,ds.
$$
Cancelling the common value in numerator and denominator we obtain
\Eq
F_1(t)
= \int_0^t S_{T_2}(s)\, dF_1(s),\qt{}
F_2(t)= \int_0^t S_{T_1}(s)\, dF_2(s).
\label{cif.simple}
\EndEq

Identity~\mref{cif.simple} provides a method for specifying the CIF in terms
of the hazard function.  A flexible choice is the lognormal hazard.
This equals the hazard for the random variable $T_j$ that is
normally distributed under a log base-e transformation,
$$
\ln T_j \sim \N(\m_j,\s_j^2) .
$$
Let $\phi_{\m,\s}$ and $\Phi_{\m,\s}$ denote the density and CDF for a
$\N(\m,\s^2)$ random variable. By~\mref{cif.simple} we have
\Array
F_1(t)
&=& \int_0^t \PP\{T_2\ge s\}\, d\PP\{T_1\le s\}\\
&=& \int_0^t \PP\{\ln T_2\ge \ln s\}\, d\PP\{T_1\le s\}\\
&=& \int_0^t \[1-\Phi_{\m_2,\s_2}(\ln s)\] d\PP\{T_1\le s\}\\
&=& \int_0^t d\PP\{T_1\le s\}
- \int_0^t \Phi_{\m_2,\s_2}(\ln s)\, d\PP\{T_1\le s\}\\
&=& \PP\{\ln T_1\le \ln t\} - \int_0^t \Phi_{\m_2,\s_2}(\ln s)\, d\PP\{\ln T_1\le \ln s\}\\
&=& \Phi_{\m_1,\s_1}(\ln t) - \int_0^t
\Phi_{\m_2,\s_2}(\ln s)\,\frac{1}{s}\,\phi_{\m_1,\s_1}(\ln s)\, ds.
\EndArray
Similarly, we have
$$
F_2(t)=\Phi_{\m_2,\s_2}(\ln t) - \int_0^t
\Phi_{\m_1,\s_1}(\ln s)\,\frac{1}{s}\,\phi_{\m_2,\s_2}(\ln s)\, ds.
$$
Both $F_1$ and $F_2$ can be rapidly computed numerically using standard
software.

Once the CIF is determined, parameters $m_d$ and $c_d$
are obtained as follows:
$$
m_d = \PP\{d-1<T\le d, \d=2|T\geq d-1\}= \frac{F_2(d)-F_2(d-1)}{S(d-1)} ,
$$
\Eq
c_d = \PP\{d-1<T\le d, \d=1|T\geq d-1\}= \frac{F_1(d)-F_1(d-1)}{S(d-1)}.
\label{mdcd}
\EndEq
Note that while the dynamic model~\mref{dynamic.model} implicitly
assumes a time window of $[0,M]$, it is not necessary to impose this
constraint in the competing risk analysis.  
This alleviates restrictive assumptions on the
survival model, but more importantly allows survival quantities to be
fully data driven.  This is especially useful when fully nonparametric
methods for estimating the CIF are utilized~\cite{ishwaran2014random}.

\section{Transmission number specification}
\label{section:para.rep}

The daily transmission numbers $R_t$ is determined by the basic reproduction
number $R_0$, the containment measures on day $t$, and the percentage
of uninfected people. It is assumed that cured cases will not get
infected again. Since $R_0$ is a
constant, we only need to set
$$
R_t=r_t \times \frac{P_{\pop}-P_t-\sum_{i=1}^{t}(D_i+C_i)}{P_{\pop}},
$$
where $D_i=\sum_{d=2}^{M}p_{i-1,d}m_d$ and
$C_i=\sum_{d=2}^{M}p_{i-1,d}c_d$ are the 
number of deaths and number of cured patients on day $t=i$
respectively, and $P_{\pop}$ denotes the total population.  The
crucial parameter is $r_t$ which is used to specify the containment
scenario. The whole model is comparable to a discrete SIR model \cite{kermack1927contribution} where $r_t$ serves as the effect contact rate and ${P_{\pop}-P_t-\sum_{i=1}^{t}(D_i+C_i)}$ serves as the susceptible population.

For initialization, values are
generated from Poisson distribution to mimic the individual variation
\cite{lloyd2005superspreading}, where
$p_{1,d}=\sum_{i=1}^{P_1}1\{X_i=d\}$,
$p^{\text{imp}}_{t,d}=\sum_{j=1}^{P^{\text{imp}}_t}1\{X_j=d\}$ and $(X_i,X_j)_{i,j}$
are independently distributed from a Poisson
distribution with mean $\l$.

\section{Results and conclusion}
\label{section:results}

To compare different pandemic scenarios, consider a region who will
experience a COVID-19 outbreak in the scenario illustrated in
Table~\ref{simulation}.  The first set of parameters are disease
related and include parameters used for the survival analysis.  For
this we use a lognormal hazard and we are comparing two treatment
plans: for scenario A and B, the mortality rate is $m_{\death}=0.04$ in
50 days, $\s_1=0.3, \s_2=0.7$, $\mu_1=3.19$ and $\mu_2=4.57$; for
scenario C, we suppose a new effective drug is available and the
mortality rate is $m_{\death}=0.02$ in 40 days, $\s_1=0.3, \s_2=0.7$,
$\mu_1=2.95$ and $\mu_2=4.6$. 
The second set of
parameters are population related.  The third parameter is $r_t$ which
defines the containment strategy.  For example, $r_t=0.21$ from
strategy A implies every 100 infected cases will communicate
to 21 individuals per day on average.  Scenario A adopts a constant
containment strategy.  Containment strategies for scenarios B and C are the same, which are adaptive and allowed to change weekly.  The averages of $r_t$ for scenario A, B
and C are all 0.21; thus all strategies have the same overall
strict level.

\begin{figure}[p]
	\centering
	\includegraphics[width=2.25in,page=2]{rplotCS}\includegraphics[width=2.25in,page=3]{rplotCS}\includegraphics[width=2.25in,page=1]{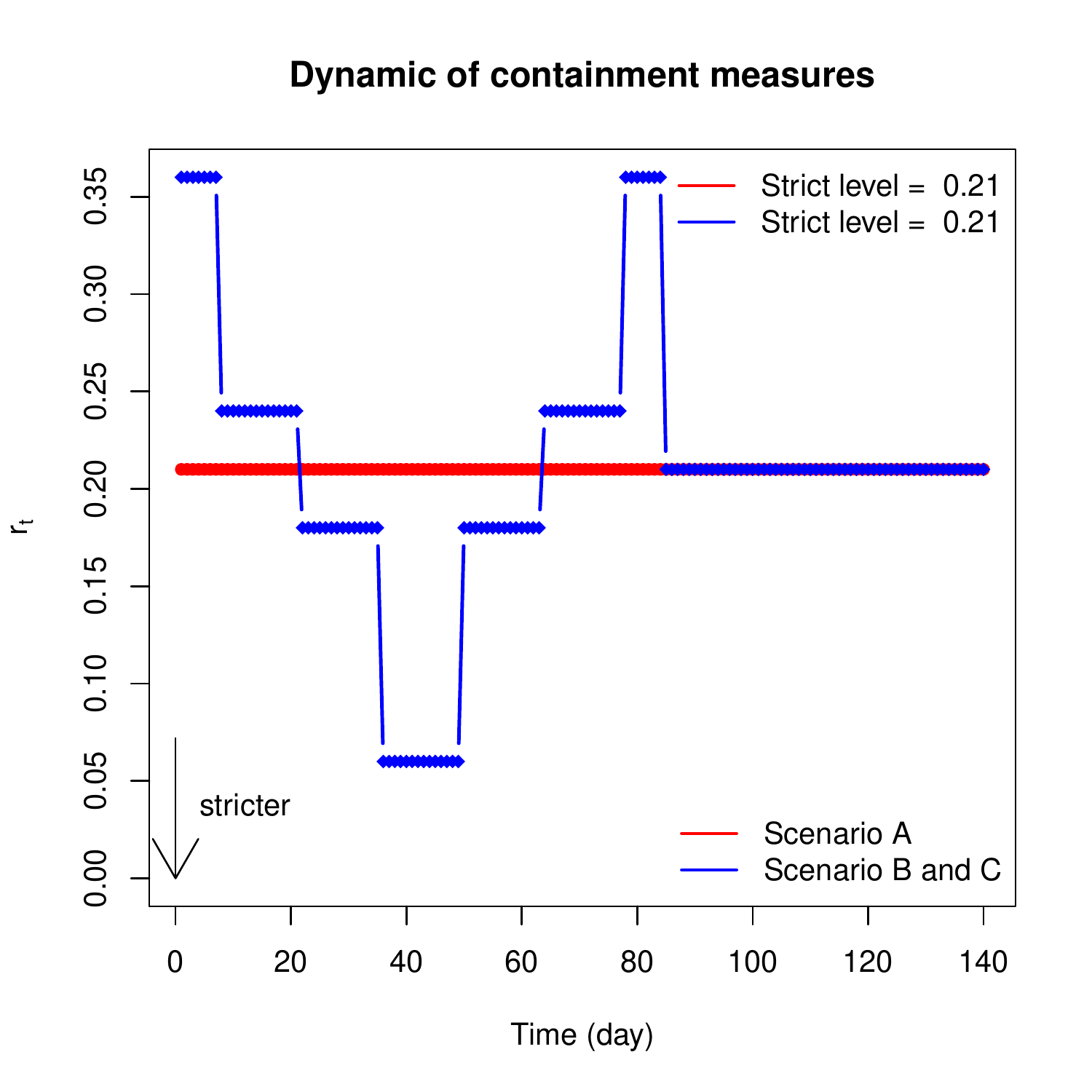}\\
	\hskip 40pt   (a)\hskip 130pt (b) \hskip 120pt (c) $\qquad$\\
	\includegraphics[width=3in,page=4]{rplotCS}\includegraphics[width=3in,page=5]{rplotCS}\\
		\hskip 40pt (d) \hskip 210pt (e) $\qquad$\\
	\caption
	{Comparison of containment strategies and treatment plans for
          disease using inputs of Table~\ref{simulation}.  Death and
          cure rate are plotted in sub-figures (a) and (b), where
          scenario A and B, colored in blue, have the same mortality
          rate and a new drug is supposed to be available in scenario
          C (colored in purple), with lower mortality rate and shorter
          infectious period. Sub-figure (c) demonstrates the different
          containment strategies across time. Scenario A (red) has a
          constant strict level while strictness level is allowed to
          change weekly for strategies B and C (blue). All containment
          measures have the same overall strict level.  From
          sub-figures (d) and (e), adaptive containment
          measures (scenario B and C) result in the smallest number
          of infected patients and deaths and end the outbreak
          faster. A new effective drug, illustrated in scenario C,
          could dramatically decrease the number of deaths and shorten
          the outbreak length.}
        \label{results}
\end{figure}

Results are displayed in Figure~\ref{results}.  After monitoring 100
simulations, the dynamic of number of infectious cases does not change
much from random initialization.  In total, numbers of deaths from
scenarios A, B and C are $7.20\times 10^3$, $5.41\times 10^3$ and
$2.49\times 10^3$; numbers of infected cases are $1.76\times 10^5$,
$1.32\times 10^5$ and $1.28\times 10^5$. The number of infectious
cases, $P_t$, reaches its peak on the 47th, 40th and 40th day and the
number of deaths, $D_t$, reaches its peak on the 60th, 52th and 49th
day for scenarios A, B and C. After the peak of $P_t$, the containment
strategy does not make much difference on the trend of $P_t$ or $D_t$.

In conclusion, compared with a constant containment strategy, adaptive
containment strategies shorten the outbreak length. Adaptive
strategies are less strict at the beginning, which results in more
severe spread. However, the stricter measures that are enforced after
this have the effect of shortening the outbreak length.  Fine tuning
these stricter adaptive measures is critical to achieving a minimum
death rate and/or reducing maximum daily number of cases. New
effective treatment is the key to death rate.  Scenario C assumes a
new treatment that reduces mortality rate within an infectious period
from 4\% to 2\%, a 50\% decrease.  When applied in our model, this
leads to a decrease in total number of deaths by 53.97\%.
Importantly, notice this value is larger than 50\% as the new
treatment reduces the number of infections due to a shorter infectious
period and cure time.

\begin{table}[htb!]
	 \caption{Necessary inputs for policy makers to compare different scenarios.\label{simulation}}
	\begin{tabular}{@{}lll@{}}
		\toprule
		Domain   & Value                & Description           \\ \midrule
		{\multirow{1}{*}{Disease}}      & 
		$M$: $M_A=M_B=50$                       &  Infected cases will be either cured or dead within $M$ days.                     \\
	                             &   	
	    $\qquad M_C=40$           & $\,\,\,\qquad$ A new effective drug is available in  scenario C            
	                              \\ 
	                              &
	    $m_{\death}=4\%\text{ or } 2\%$          &    Within $M$ days, $m_{\death}$ of infected cases will be dead.      \\ &
	    $\sigma_1=0.3,\mu_1=3.19\text{ or }2.95$         &    Parameters to shape the cure hazard function.      \\  &
	    $\sigma_2=0.7,\mu_2=4.57\text{ or }4.6$         &    Parameters to shape the death hazard function.      \\ \midrule
		{\multirow{1}{*}{People}}     & 
		$P_{\pop}=2.8\times 10^6$              &   On day 1,
                $P_{\pop}$ people are not infected within the region.                    \\
		                                & 
		$P_1=200$              &   On day 1, $P_1$ individuals are infectious.\\
		                              &
		$P^{\imp}_{15}=P^{\imp}_{48}=2$                       &        On days 15, 29, 48 and 63, there are 2, 4, 2 and 4               \\ 
                             &                      
        $P^{\imp}_{29}=P^{\imp}_{63}=4$                       &        $\,\,\,\qquad$infectious people who travel into the region.             \\
                              &
        $\lambda=16$                       &   Initial infectious
                cases ($P_1$ and $P^{\imp}_1$) have been              \\ 
                             &
                        &    $\,\,\,\qquad$ infected   for $\l$ days
                on average.                 \\ \midrule
		{\multirow{1}{*}{Policy}} & 
		$r_t$ described in Figure 1(c)                       &       Smaller value represent stricter containment measures*.\\
                              \midrule
	\end{tabular}

        *$r_t$ can be interpreted as the average number of newly
        infected case communicated \textit{per infectious person per
          day} on day $t$, if nearly all the population is
        uninfected. The model will adjust these inputs with percentage
        of infected cases across time, which produces $R_t$.   
        
\end{table}

\section*{Supplement}
An online prediction tool is available at$\,$    \url{https://minlu.shinyapps.io/killCOVID19/}.
\vskip0.4in
\bibliographystyle{abbrv}

\bibliography{cv19}

\end{document}